\documentclass[12pt]{article}
\usepackage{amssymb,amsmath,epsfig}
\allowdisplaybreaks
\begin{document}
\title{\bf Aspects of Some New Versions of Pilgrim Dark Energy in DGP Braneworld}

\author{Abdul Jawad$^1$ \thanks {abduljawad@ciitlahore.edu.pk,
jawadab181@yahoo.com}, Shamaila Rani$^1$
\thanks{shamailatoor.math@yahoo.com;
drshamailarani@ciitlahore.edu.pk} , Ines G. Salako$^{2,3}$
\thanks{ines.salako@imsp-uac.org,
inessalako@gmail.com}\\ and Faiza Gulshan$^4$
\thanks{fazi.gull@yahoo.com}
\\$^1$Department of Mathematics, COMSATS Institute of\\ Information
Technology, Lahore, Pakistan.\\
$^2$Institut de Math\'ematiques et de Sciences Physiques\\ (IMSP) 01
BP 613 Porto-Novo, B\'enin\\
$^3$D\'epartement de Physique - Universit\'e d'Agriculture\\ de
K\'etou BP 13 K\'etou, B\'enin\\
$^4$ Department of Mathematics, Lahore Leads university,\\
Lahore-54590, Pakistan.}
\date{}

\maketitle
\begin{abstract}

The illustration of cosmic acceleration under two interacting dark
energy models (pilgrim dark energy with Granda and Oliveros cutoff
and its generalized ghost version) in DGP braneworld framework is
presented. In the current scenario, the equation of state parameter,
deceleration parameter, $\omega_{D}-\omega'_{D}$ plane and
statefinder diagnosis are investigated. The equation state parameter
behave-like phantom era of the universe. The deceleration parameter
depicts the accelerated expansion of the universe in both models.
The cosmological planes like $\omega_{D}-\omega'_{D}$ and
statefinder corresponds to $\Lambda$CDM limit. To end, we remark
that our results support to phenomenon of pilgrim dark energy and
cosmic acceleration. Also, the results are consistent with
observational data.
\end{abstract}

\section{Introduction}

It has been confirmed by current observational data that the our
universe undergoes accelerated expansion \cite{auna}-\cite{adeuxb}.
It is consensus that this accelerated expansion phenomenon is due to
a mysterious form of force called dark energy (DE). This can be
explained through the well-known parameter called equation of state
(EoS) parameter $\omega_{D}$. It is suggested through WMAP data that
the value of EoS parameter is bounded as $-1.11<\omega_{D}<-0.86$
\cite{d1}. This could be consistent if DE behaves like cosmological
constant with $\omega_{D}=-1$ and therefore our universe seems to
approach asymptotically a de Sitter universe.

In order to describe accelerated expansion phenomenon, two different
approaches has been adopted. One is the proposal of various
dynamical DE models such as family of chaplygin gas \cite{2s},
holographic \cite{3s,4s}, new agegraphic \cite{5s}, polytropic gas
\cite{6s}, pilgrim \cite{N5}-\cite{N7} $DE$ models etc. A second
approach for understanding this strange component of the universe is
modifying the standard theories of gravity, namely, General
Relativity (GR) or Teleparallel Theory Equivalent to GR (TEGR).
Several modified theories of gravity are $f(R)$, $f(T)$
\cite{st1}-\cite{st6}, $f(R,\mathcal{T})$ \cite{ma1}-\cite{ma2},
$f(G)$ \cite{mj1}-\cite{mj5} (where $R$ is the curvature scalar, $T$
denotes the torsion scalar, $\mathcal{T}$ is the trace of the energy
momentum tensor and $G$ is the invariant of Gauss-Bonnet defined as
$G=R^2-4R_{\mu\nu}R^{\mu\nu}+ R_{\mu\nu
\lambda\sigma}R^{\mu\nu\lambda\sigma}$).

Special attention is attached to the so-called braneworld model
proposed by Dvali, Gabadadze, and Porrati (DGP)
\cite{DGP}-\cite{Deffayet1} (for reviews, see also \cite{Koyama}).
In a cosmological scenario, this approach leads to a late-time
acceleration as a result of the gravitational leakage from a
$3$-dimensional surface ($3$-brane) to a $5$-th extra dimension on
Hubble distances. Hirano and Komiya~\cite{Hirano} have generalized
the modified Friedmann equation as suggested by Dvali and
Turner~\cite{Dvali} for achieving the phantom-like gap with an
effective energy density with an EoS with $\omega_{D}<-1$.

The DGP model presents two branches of solutions, i.e., the
self-accelerating branch ($\epsilon=+1$) and the normal
($\epsilon=-1$) one. The self accelerating branch leads to an
accelerating universe without using any exotic fluid, but shows
problems like ghost~\cite{Koyama1}. However, the normal branch need
a DE component which is compatible with the observational
data~\cite{Lue,Lazkoz}. The extension of these models have been
studied in \cite{Mariam} for $f(R)$ gravity in order to obtain a
self acceleration. The attempts of solutions for a DGP brane-world
cosmology with a k-essence field were found in \cite{MariamChimento}
showing big rip scenarios and asymptotically de Sitter phase in the
future. DGP model has also been discussed by various observations
without DE model (Self-accelerating DGP branch) \cite{28}-\cite{31}
and with DE model (normal branch) \cite{DGP1}-\cite{DGP3}. However,
in normal branch, the addition of dynamical DE model provides us new
different structures to describe the late time acceleration with
better cosmological solutions on the brane.

Holographic DE (HDE) model is the most prominent for interpreting
the DE scenario and its idea comes from the unification attempt of
quantum mechanics and gravity. According to t' Hooft, quantum
gravity demands three-dimensional world as a holographic image (an
image whose all data can be stored on a two-dimensional projection).
He stated it in the form of holographic principle, i.e., \emph{all
the information relevant to a physical system inside a spatial
region can be observed on its boundary instead of its volume}
\cite{RR14}. The construction of HDE density is based on Cohen et
al. \cite{RR15} relation about the vacuum energy of a system with
specific size whose maximum amount should not exceed the BH mass
with the same size. This can be expressed as
$L^{3}\rho_{\Lambda}\leq L M^{2}_{p}$, where $M^{2}_{p}=(8 \pi
G)^{-1}$ is the reduced Planck mass and $L$ represents the IR
cutoff. We can get HDE density from the above inequality as
\begin{equation}\label{0.2.1}
\rho_{\Lambda}=3c^{2}M^{2}_pL^{-2},
\end{equation}
here $c$ is the dimensionless HDE constant parameter. The
interesting feature of HDE density is that it provides a relation
between ultraviolet (bound of vacuum energy density) and IR (size of
the universe) cutoffs. However, a controversy about the selection of
IR cutoff of HDE has been raised since its birth. As a result,
different people have suggested different expressions.

In the present paper, we check the role of some new models of
pilgrim DE (PDE) (pilgrim dark energy with Granda and Oliveros (GO)
cutoff and its generalized ghost version) in DGP Braneworld. We
develop different cosmological parameters and planes. This paper is
outlined as follows: In section \textbf{2}, we provide the basics of
the DGP braneworld model and explain the PDE models. Sections
\textbf{3} and \textbf{4} are devoted for cosmological parameters
and cosmological planes for new models of PDE. In the last section,
we conclude our results.

\section{DGP Braneworld Model and Pilgrim Dark Energy}

Now we define the cosmological evolution on the brane by Friedmann
equation as \cite{Koyama}
\begin{equation}\label{2}
H^{2}+\frac{k}{a^{2}}=\bigg(\sqrt{\frac{\rho}{3}+\frac{1}{4r_{c}^{2}}}+
\frac{\epsilon}{2r_{c}}\bigg)^{2},
\end{equation}
where $\rho=\rho_{M}+\rho_{D}$ is the total cosmic fluid energy
density on the brane ($\rho_{M}$ is the CDM density while $\rho_{D}$
is the PDE density). Also, $r_{c}$ is the crossover length given by
\cite{20}
\begin{equation}\label{3}
r_{c}=\frac{M_{pl}^{2}}{2M_{5}^{3}}=\frac{G_{5}}{2G_{4}},
\end{equation}
$r_{c}$ is defined as a distance scale reflecting the competition
between $4D$ and $5D$ effects of gravity. Below the length $r_{c}$,
gravity appears $4$-dim and above the length $r_{c}$, gravity can
leak into the extra dimension. 
For the spatially flat DGP Braneworld
$(k=0)$, the Friedmann equation (\ref{3}) reduces to
\begin{equation}\label{4}
H^{2}-\frac{\epsilon}{r_{c}}H=\frac{1}{3}(\rho_{M}+\rho_{D}).
\end{equation}
Since, we are taking the interaction between DE and CDM, hence the
conservation equations turn out to be
\begin{equation}\label{5}
\dot{\rho}_{M}+3H\rho_{M}=Q\quad
\dot{\rho}_{D}+3H(\rho_{D}+p_{D})=-Q,
\end{equation}
here $Q$ describes the interaction between PDE and CDM. We choose
$Q=3b^{2}H\rho_{m}$ as an interaction term  with $b^{2}$ being a
coupling constant. This interaction term is used for transferring
the energy through different cosmological constraints. Its positive
sign indicates that DE decays into CDM and negative sign shows that
CDM decays into DE. Here, we take $Q$ as positive because it is more
favorable with observational data. Hence, with this interaction
form, Eq. (\ref{5}) provides
\begin{equation}\label{7}
\rho_{M}=\rho_{M_{0}}a^{3(b^{2}-1)},
\end{equation}
here $\rho_{M_{0}}$ is the integration constant.

Further, we illustrate the discussion about under consideration
model called PDE. Cohen et al. \cite{RR14} relation leads to the
bound of energy density from the idea of formation of BH in quantum
gravity. However, it is suggested that formation of BH can be
avoided through appropriate repulsive force which resists the matter
collapse phenomenon. The phantom-like DE possesses the appropriate
repulsive force (in spite of other phases of DE like vacuum or
quintessence DE). By keeping in mind this phenomenon, Wei \cite{N5}
has suggested the DE model called PDE on the speculation that
phantom DE possesses the large negative pressure as compared to the
quintessence DE which helps in violating the null energy condition
and possibly prevent the formation of BH. In the past, many
applications of phantom DE exist in the literature. It is also
playing an important role in the reduction of mass due to its
accretion process onto BH. Many works have been done in this support
through a family of chaplygin gas \cite{30a}-\cite{30d}. For
instance, phantom DE is also play an important role in the wormhole
physics where the event horizon can be avoided due to its presence
\cite{shara}-\cite{shard}.

It was also argued that BH area reduces up to $50$ percent through
phantom scalar field accretion onto it \cite{9}. According to Sun
\cite{10}, mass of BH tends to zero when the universe approaches to
big rip singularity. It was also suggested that BHs might not be
exist in the universe in the presence of quintessence-like DE which
violates only strong energy condition \cite{11}. However, these
works do not correspond to reality because quintessence DE does not
contain enough resistive force to in order to avoid the formation of
BH. Also, Saridakis et al. \cite{S1}-\cite{S9} have obtained the
phantom crossing, quintom as well as phantom-like nature of the
universe in different frameworks and found interesting results in
this respect.

The above discussion is motivated to Wei \cite{N5} in developing the
$PDE$ model. He analyzed this model with Hubble horizon through
different theoretical as well as observational aspects. The energy
density for PDE model with hubble horizon is defined as
\begin{equation}\label{8}
\rho_{D}=3c^{2}L^{-u}.
\end{equation}
where $u$ is the PDE parameter, $m_{pl}=1$ and $L$ is known as IR
cutoff. The proposal of PDE model by Wei \cite{N5} is based on two
properties. The first property of PDE is
\begin{eqnarray}\label{pde}
\rho_{D}\gtrsim m_p^2 L^{-2}.
\end{eqnarray}
From Eqs.(\ref{8}) and (\ref{pde}), we have $L^{2-u}\gtrsim
m_p^{u-2}=l_p^{2-u}$, where $l_p$ is the reduced Plank length. Since
$L>l_p$, one requires
\begin{eqnarray}\label{sless2}
u\leq 2.
\end{eqnarray}
The second requirement for PDE is that it gives phantom-like
behavior \cite{N5}
\begin{eqnarray}\label{wpde}
\omega_{D}<-1
\end{eqnarray}
It is stated \cite{N5} that a particular cutoff $L$ has to choose to
obtain the EoS for PDE. For instance, radius of Hubble horizon
$L=H^{-1}$, event horizon $L=R_{E}=a\int_{t}^{\infty}\frac{dt}{a}$,
the form $L=(H^2+\dot{H})^{-\frac{1}{2}}$ represented the Ricci
length, the GO length $(\alpha H^2+\beta \dot{H})^{-\frac{1}{2}}$,
etc.

Recently, we have investigated this model by taking different IR
cutoffs in flat as well as non-flat FRW universe with different
cosmological parameters as well as cosmological planes \cite{N6,N7}.
This model has also been investigated in different modify gravity
theories \cite{N47}-\cite{N49}. In the next two sections, we will
discuss the cosmological parameters of PDE with GO cutoff and ghost
version of PDE.

\section{Pilgrim Dark Energy with Granda and Oliveros Cutoff}

\textbf{Granda and Oliveros \cite{RR17} deveoped IR cutoff
(involving Hubble parameter and its derivative) of HDE parameterized
by two dimensionless constants called new HDE (NHDE). They suggested
that this model can be an effective candidate in solving the cosmic
coincidence problem. Yu et al. \cite{RR35} analyzed the behavior of
interacting NHDE with CDM and found that this model inherits the
features of already presented HDE models. Also, constraints on
different cosmological parameters are established for this model by
using the data of different observational schemes and Markov chain
Monte Carlo method \cite{RR36}.} The GO cutoff can be defined as
follows \cite{46}
\begin{equation}\label{18}
\rho_{D}=3(\alpha H^{2}+\beta\dot H).
\end{equation}
Where $\alpha$ and $\beta$ are the positive constants. With GO
cutoff, the PDE model turns out to be
\begin{equation}\label{19}
\rho_{D}=3(\alpha{H^{2}}+\beta{\dot{H}})^{u/2}.
\end{equation}
By taking the time derivative of Eq.(\ref{4}) and then by using the
Eq.(\ref{19}), we get
\begin{equation}\label{20}
\frac{\dot{H}}{H^{2}}=\frac{1}{H^{2}\beta}\bigg(H^{2}(1-2\epsilon\sqrt{\Omega_{r_{c}}})
-\frac{1}{3}\rho_{m_{_{0}}}a^{-3(1-b^{2})}\bigg)^{2/u}-
\frac{\alpha}{\beta}.
\end{equation}
where $\Omega_{r_{c}}=\frac{1}{4H^2r^2_c}$. By solving the
Eqs.(\ref{5}) and (\ref{19}), we obtain the EoS parameter
\begin{eqnarray}\nonumber
\omega_{D}&=&-1-\frac{2(1-\epsilon\sqrt{\Omega_{r_{c}}})}{3H^{2}}\big(
(1-2\epsilon\sqrt{\Omega_{r_{c}}})-\frac{\rho_{m_{_{0}}}a^{-3(1-b^{2})}}{3H^{2}}\big)^{-1}
\\\nonumber&\times&\frac{1}{\beta}\bigg(\big(H^{2}(1-2\epsilon\sqrt{\Omega_{r_{c}}})-\frac{1}{3}
\rho_{m_{_{0}}}a^{-3(1-b^{2})}\big)^{2/u}-\alpha
H^{2}\bigg)\\\label{21}&+&\frac{(2b^{2}+1)\rho_{m_{_{0}}}a^{-3(1-b^{2})}}{3H^{2}}\bigg(
(1-2\epsilon\sqrt{\Omega_{r_{c}}})-\frac{\rho_{m_{_{0}}}a^{-3(1-b^{2})}}{3H^{2}}\bigg)^{-1}.
\end{eqnarray}
\begin{figure} \centering
\epsfig{file=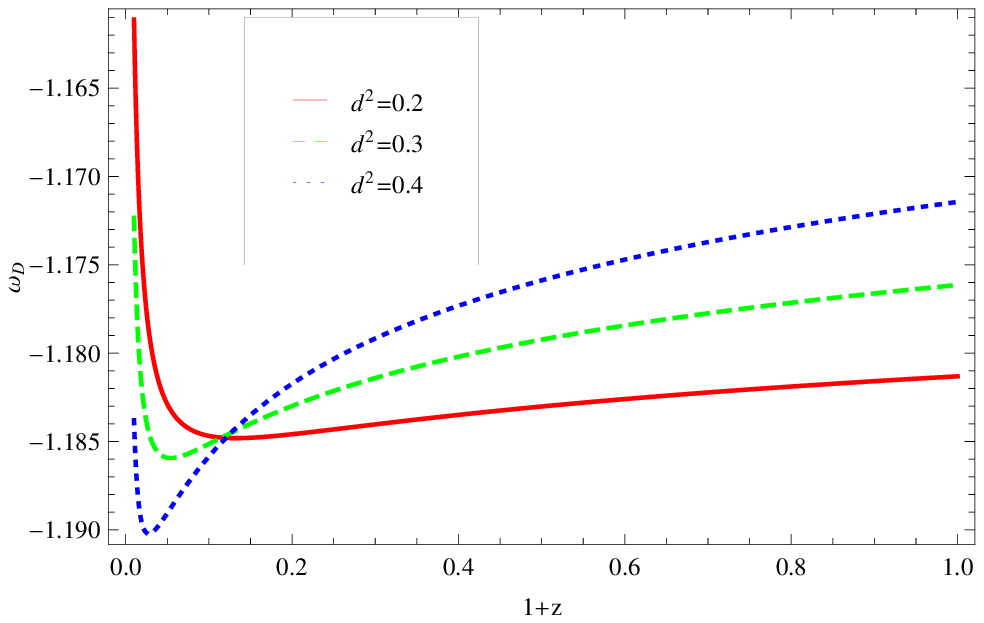,width=.50\linewidth}\epsfig{file=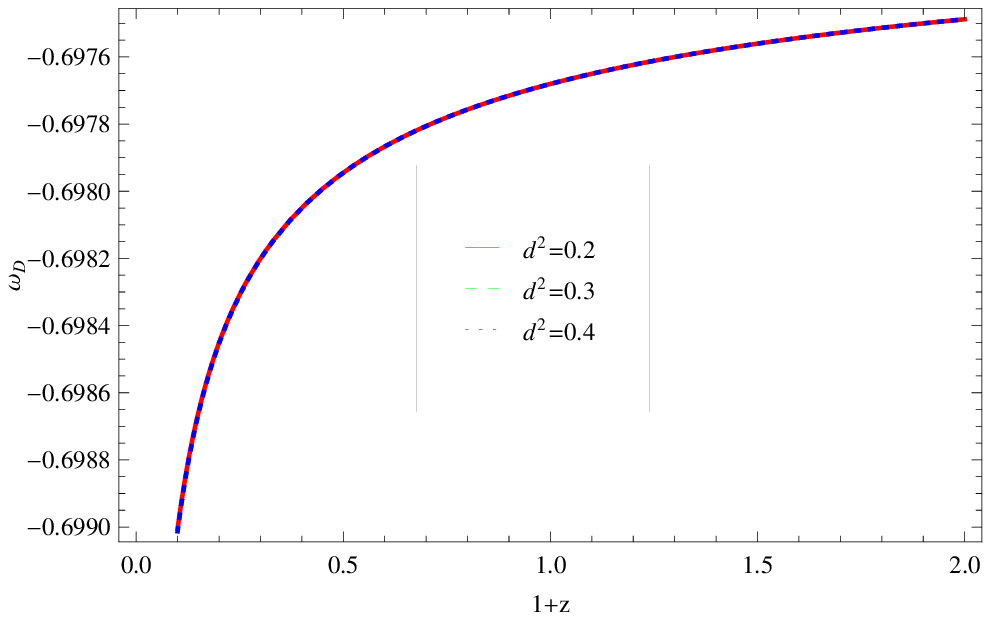,width=.50\linewidth}\\
\epsfig{file=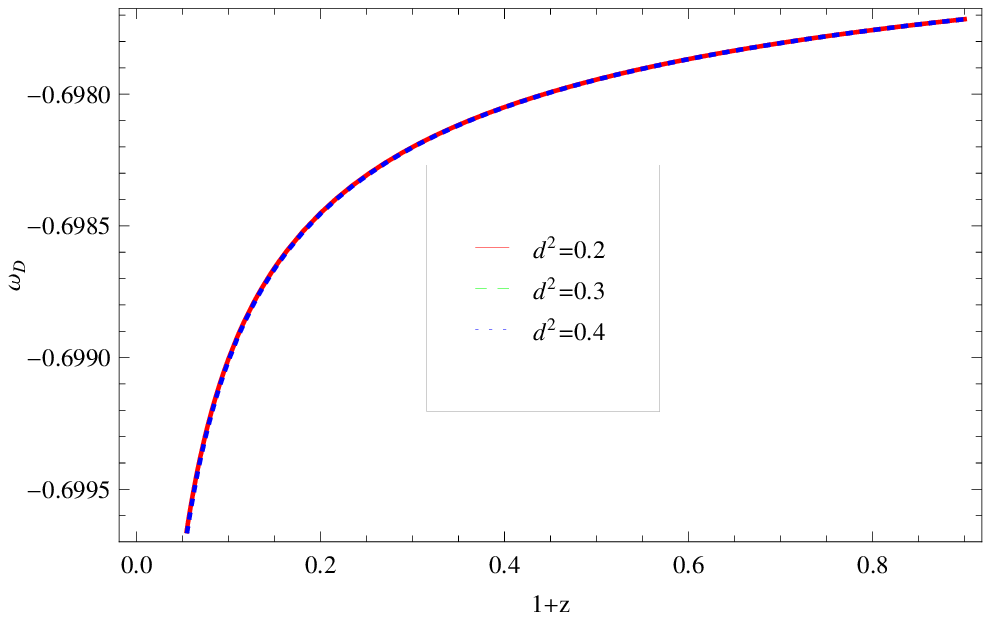,width=.50\linewidth}\caption{Plots
of $\omega_{D}$ versus $1+z$ for new PDE model in DGP with u=1
(upper left panel), $u=-1$ (upper right panel), $u=-2$ (lower
panel), respectively.}
\end{figure}

The deceleration parameter can be defined as follows:
\begin{equation}\label{22a}
q=-1-\frac{\dot{H}}{H^2}.
\end{equation}
The deceleration parameter can also be obtain by using the Eqs.
(\ref{20}) and (\ref{22a}) as
\begin{equation}\label{22}
q=-1-\frac{1}{\beta
H^{2}}\bigg(H^{2}(1-2\epsilon\sqrt{\Omega_{r_{c}}})-\frac{1}{3}
\rho_{m_{_{0}}}a^{-3(1-b^{2})}\bigg)^{2/u}+\frac{\alpha}{\beta}.
\end{equation}
\begin{figure} \centering
\epsfig{file=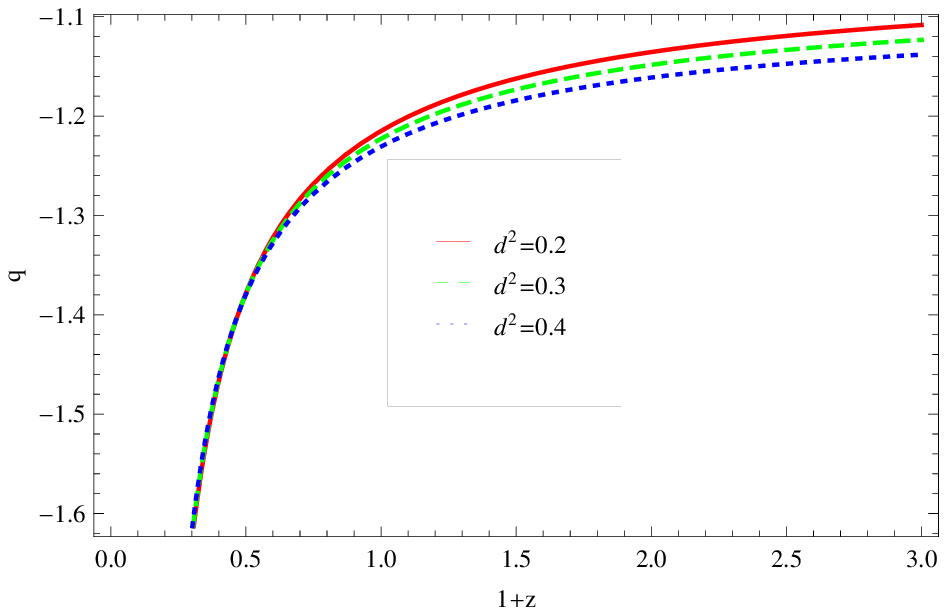,width=.50\linewidth}\epsfig{file=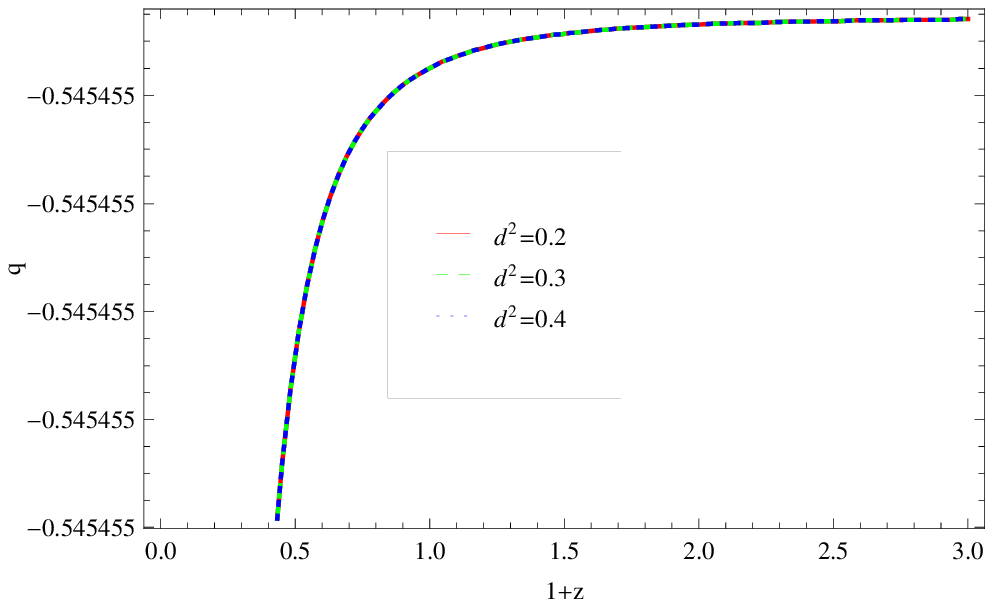,width=.50\linewidth}\\
\epsfig{file=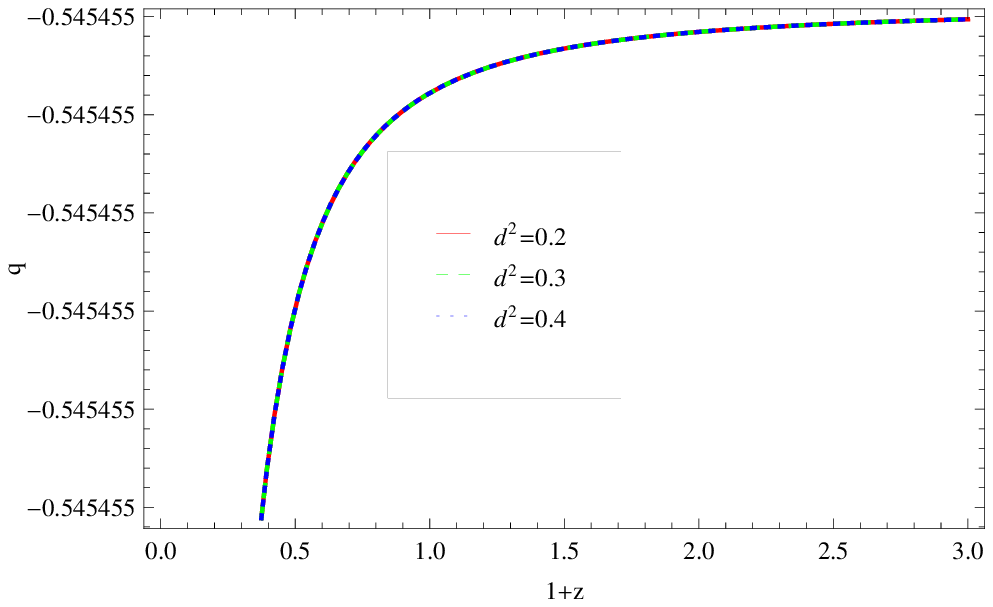,width=.50\linewidth}\caption{Plots
of $q$ versus $1+z$ for new PDE model in DGP with u=1 (upper left
panel), $u=-1$ (upper right panel), $u=-2$ (lower panel),
respectively.}
\end{figure}

\textbf{Here, we find the regions on the
$\omega_{\vartheta}-\omega'_{\vartheta}$ plane
($\omega'_{\vartheta}$ represents the evolution of
$\omega_{\vartheta}$) as defined by Caldwell and Linder \cite{75}
for models under consideration. The models can be categorized in two
different classes as thawing and freezing regions on the
$\omega_{\vartheta}-\omega'_{\vartheta}$ plane. The thawing models
describe the region $\omega'_{\vartheta}>0$ when
$\omega_{\vartheta}<0$ and freezing models represent the region
$\omega'_{\vartheta}<0$ when $\omega_{\vartheta}<0$. Initially, this
phenomenon was applied for analyzing the behavior of quintessence
model and found that the corresponding area occupied on the
$\omega_{\vartheta}-\omega'_{\vartheta}$ plane describes the thawing
and freezing regions. }

The differentiation of EoS parameter Eq. (\ref{21}) w.r.t. $x=\ln a$
leads to
\begin{eqnarray}\nonumber
\omega'_{D}&=&-\frac{1}{3}\bigg(H^{2}(1-2\epsilon\sqrt{\Omega_{r_{c}}})-\frac{1}{3}
\rho_{m_{_{0}}}a^{-3(1-b^{2})}\bigg)^{-1}\bigg(\frac{2}{H}(1-\epsilon\sqrt{\Omega_{r_{c}}})
\\\nonumber&\times&\bigg(\frac{2}{u\beta}\big(H^{2}(1-2\epsilon\sqrt{\Omega_{r_{c}}})-\frac{1}{3}
\rho_{m_{_{0}}}a^{-3(1-b^{2})}\big)^{\frac{2}{u}-1}\big(2(1-\epsilon\sqrt{\Omega_{r_{c}}})
\\\nonumber&\times&\big(\frac{1}{\beta
}\big(H^{2}(1-2\epsilon\sqrt{\Omega_{r_{c}}})-\frac{1}{3}
\rho_{m_{_{0}}}a^{-3(1-b^{2})}\big)^{2/u}-\frac{\alpha}{\beta}H^{2}\big)+(1-b^{2})
\\\nonumber&\times&\rho_{m_{_{0}}}a^{-3(1-b^{2})}H\big)-\frac{2\alpha}{\beta}H\big(\frac{1}{\beta
}\big(H^{2}(1-2\epsilon\sqrt{\Omega_{r_{c}}})-\frac{1}{3}
\rho_{m_{_{0}}}a^{-3(1-b^{2})}\big)^{2/u}\\\nonumber&-&\frac{\alpha}{\beta}H^{2}\big)\bigg)
+2\epsilon\sqrt{\Omega_{r_{c}}}\big(\frac{1}{\beta
H^{2}}\big(H^{2}(1-2\epsilon\sqrt{\Omega_{r_{c}}})-\frac{1}{3}
\rho_{m_{_{0}}}a^{-3(1-b^{2})}\big)^{2/u}\\\nonumber&-&\frac{\alpha}{\beta}\big)-
3(1-b^{2})^{2}\rho_{m_{_{0}}}a^{-3(1-b^{2})}-9b^{2}(1-b^{2})\rho_{m_{_{0}}}a^{-3(1-b^{2})}\bigg)
+\frac{1}{3}\\\nonumber&\times&\bigg(H^{2}(1-2\epsilon\sqrt{\Omega_{r_{c}}})-\frac{1}{3}
\rho_{m_{_{0}}}a^{-3(1-b^{2})}\bigg)^{-2}\bigg(2(1-\epsilon\sqrt{\Omega_{r_{c}}})
\big(\frac{1}{\beta
}\big(H^{2}\\\nonumber&\times&(1-2\epsilon\sqrt{\Omega_{r_{c}}})-\frac{1}{3}
\rho_{m_{_{0}}}a^{-3(1-b^{2})}\big)^{2/u}-\frac{\alpha}{\beta}H^{2}\big)+(1-b^{2})
\rho_{m_{_{0}}}\\\nonumber&\times&a^{-3(1-b^{2})}+3b^{2}\rho_{m_{_{0}}}a^{-3(1-b^{2})}\bigg)
\bigg(2(1-\epsilon\sqrt{\Omega_{r_{c}}})\big(\frac{1}{\beta
}\big(H^{2}(1-2\epsilon\\\label{23}&\times&\sqrt{\Omega_{r_{c}}})-\frac{1}{3}
\rho_{m_{_{0}}}a^{-3(1-b^{2})}\big)^{2/u}-\frac{\alpha}{\beta}H^{2}\big)+(1-b^{2})\rho_{m_{_{0}}}a^{-3(1-b^{2})}\bigg).
\end{eqnarray}
\begin{figure} \centering
\epsfig{file=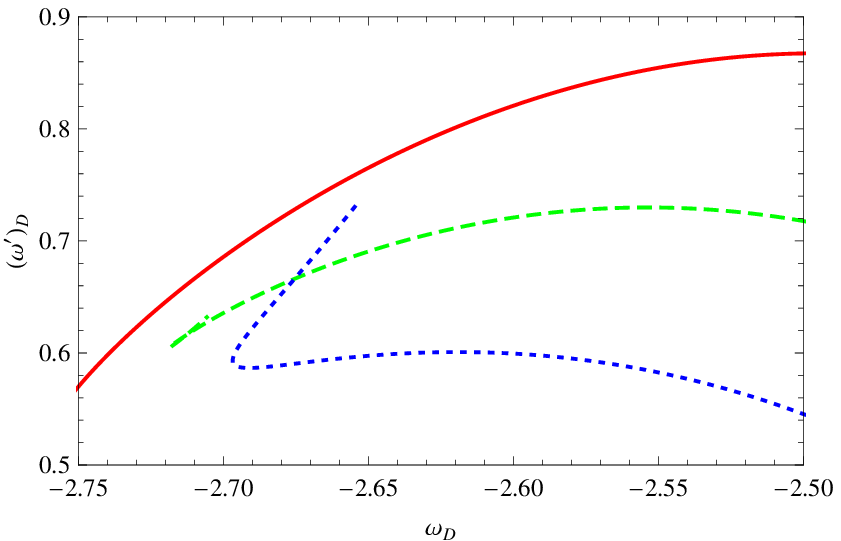,width=.50\linewidth}\epsfig{file=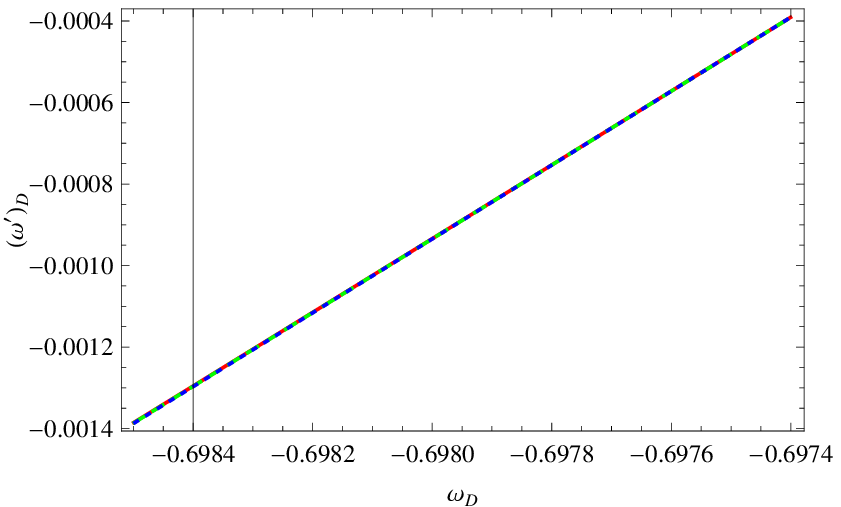,width=.50\linewidth}
\\\epsfig{file=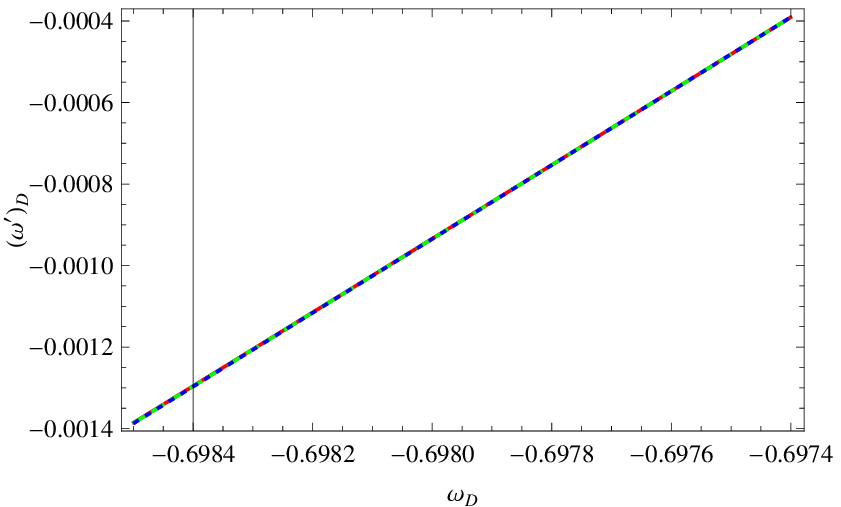,width=.50\linewidth}\caption{Plots
of $\omega_{D}-\omega'_{D}$ for new PDE model in DGP with u=1 (upper
left panel), $u=-1$ (upper right panel), $u=-2$ (lower panel),
respectively.}
\end{figure}
The Hubble parameter $H$ and the deceleration parameter $q$ cannot
discriminate among various DE models. For this purpose Sahni et al.
\cite{40} and Alam et al. \cite{41} proposed a new geometrical
diagnostic pair for DE and it is constructed from the scale factor
$a(t)$ and its derivatives up to third order. The statefinder pair
$(r,s)$ is defined as
\begin{equation}\label{14}
r=\frac{\dddot{a}}{aH^{3}},    s=\frac{r-1}{3(q-1/2)}.
\end{equation}
The state-finder parameter $r$ can be expressed as
\begin{equation}\label{15}
r=1+3\frac{\dot{H}}{H^{2}}+\frac{\ddot{H}}{H^{3}}.
\end{equation}
By using the relations (\ref{14}), (\ref{15}) and (\ref{20}), we
obtain the state-finder parameters $(r,s)$ as
\begin{eqnarray}\nonumber
r&=&1+\bigg(\frac{1}{\beta
H^{2}}\big(H^{2}(1-2\epsilon\sqrt{\Omega_{r_{c}}})-\frac{1}{3}
\rho_{m_{_{0}}}a^{-3(1-b^{2})}\big)^{2/u}-\frac{\alpha}{\beta}\bigg)\bigg(3\\\nonumber&+&\frac{4}{\beta
u}(1-\epsilon\sqrt{\Omega_{r_{c}}})(H^{2}(1-2\epsilon\sqrt{\Omega_{r_{c}}})
-\frac{\rho_{m_{_{0}}}a^{-3(1-b^{2})}}{3})^{\frac{2}{u}-1}-2
\frac{\alpha}{\beta}\bigg)\\\label{24}&+&\frac{2(1-b^{2})}{u\beta
H^{2}}\rho_{m_{_{0}}}a^{-3(1-b^{2})}\big(H^{2}(1-2\epsilon\sqrt{\Omega_{r_{c}}})-\frac{1}{3}
\rho_{m_{_{0}}}a^{-3(1-b^{2})}\big)^{\frac{2}{u}-1}. \\\nonumber
s&=&\frac{1}{3}\bigg(\frac{-3}{2}-\frac{1}{\beta
H^{2}}(H^{2}(1-2\epsilon\sqrt{\Omega_{r_{c}}}))^{2/u}+\frac{\alpha}{\beta}\bigg)^{-1}\bigg(\big(\frac{1}{\beta
H^{2}}(H^{2}\\\nonumber&\times&(1-2\epsilon\sqrt{\Omega_{r_{c}}})-\frac{1}{3}
\rho_{m_{_{0}}}a^{-3(1-b^{2})})^{2/u}+\frac{\alpha}{\beta}\big)\big(3+\frac{4}{\beta
u}(1-\epsilon\sqrt{\Omega_{r_{c}}})\\\nonumber&\times&(H^{2}(1-2\epsilon\sqrt{\Omega_{r_{c}}})
-\frac{\rho_{m_{_{0}}}a^{-3(1-b^{2})}}{3})^{\frac{2}{u}-1}-2
\frac{\alpha}{\beta}\big)+\frac{2(1-b^{2})}{u\beta
H^{2}}\\\label{25}&\times&\rho_{m_{_{0}}}a^{-3(1-b^{2})}\big(H^{2}(1-2\epsilon\sqrt{\Omega_{r_{c}}})-\frac{1}{3}
\rho_{m_{_{0}}}a^{-3(1-b^{2})}\big)^{\frac{2}{u}-1}\bigg).
\end{eqnarray}
\begin{figure} \centering
\epsfig{file=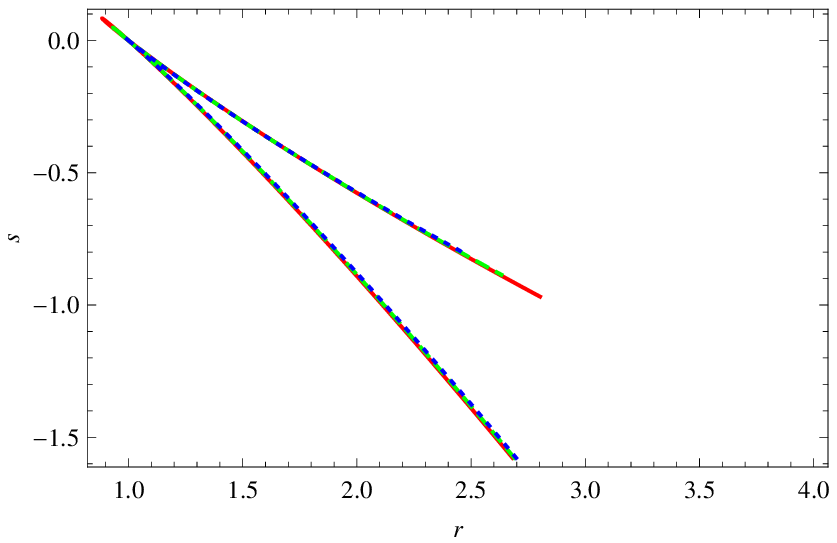,width=.50\linewidth}\epsfig{file=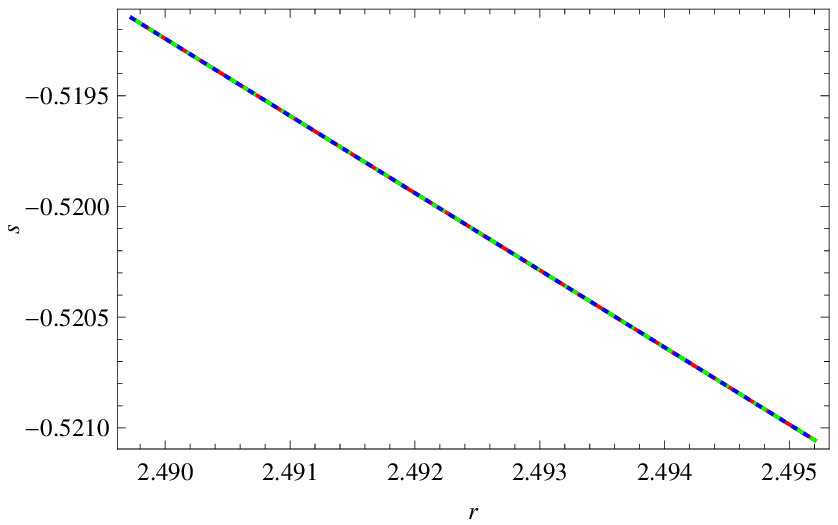,width=.50\linewidth}
\\\epsfig{file=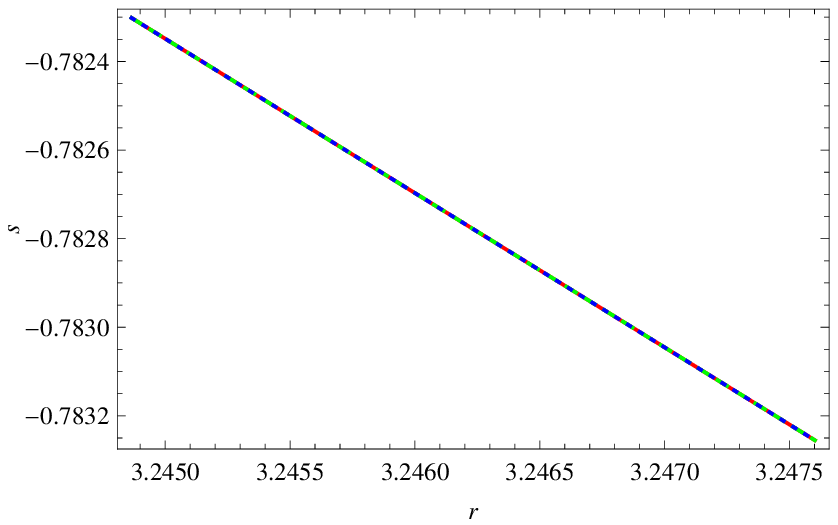,width=.50\linewidth}\caption{Plots
of $r-s$ for new PDE model in DGP with u=1 (upper left panel),
$u=-1$ (upper right panel), $u=-2$ (lower panel), respectively.}
\end{figure}
The behavior of EoS parameter in terms of redshift parameter (by
utilizing $a=(1+z)^{-1}$) is displayed in Figure \textbf{1} for
$u=1, -1, -2$ (since there is property of PDE for attaining useful
results, we should choose $u\leq2$, that is why we have chosen $u=1,
-1, -2$). The EoS parameter shows the phantom-like behavior for
$u=1$ (left panel of Figure \textbf{1}) while exhibits
quintessence-like behavior for $u=-1,-2$ (right and lower panels of
Figure \textbf{1}). However, the deceleration parameter remains less
than $0$ i.e., $q<0$ for all values of $u$ (Figure \textbf{2}).
Hence, the deceleration parameter exhibits the accelerated expansion
of the universe. We also developed $\omega_{D}-\omega'_{D}$ plane
for $u=1,-1,-2$ as shown in Figure \textbf{3}. It can be seen that
this plane corresponding to for $u=1$ lies in the thawing region,
while lie in the freezing region for $u=-1, -2$. The $r-s$ planes
for $u=1, -1, -2$ are shown in Figure \textbf{4} which behave like
chaplygin gas model. However, the $\Lambda$CDM limit is also
achieved for $u=-1$ case.

\section{Generalized Ghost Pilgrim Dark Energy}

The generalized ghost version of PDE can be defined as \cite{62}
\begin{equation}\label{28}
\rho_{D}=(\alpha H+\beta H^{2})^{u}.
\end{equation}
Equations (\ref{4}) and (\ref{28}) gives the Hubble parameter for
this model as follows
\begin{equation}\label{29}
\frac{\dot{H}}{H^{2}}=-\frac{1}{H^{2}}\bigg(\frac{(1-b^{2})\rho_{m_{0}}a^{-3(1-b^
{2})}}{2-2\epsilon\sqrt{\Omega_{r}}-\frac{u}{3H}(\alpha{H}+\beta{H^
{2}})^{u-1}(\alpha+2\beta{H})}\bigg).
\end{equation}
This leads to deceleration parameter as follows
\begin{equation}\label{30}
q=-1+\frac{1}{H^{2}}\bigg(\frac{(1-b^{2})\rho_{m_{0}}a^{-3(1-b^
{2})}}{2-2\epsilon\sqrt{\Omega_{r}}-\frac{u}{3H}(\alpha{H}+\beta{H^
{2}})^{u-1}(\alpha+2\beta{H})}\bigg).
\end{equation}
\begin{figure} \centering
\epsfig{file=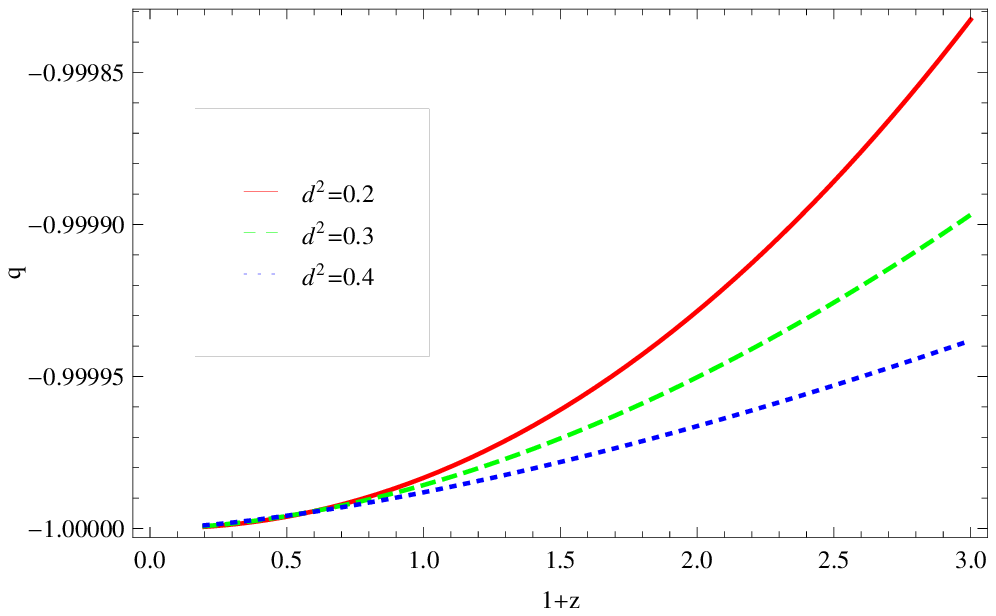,width=.50\linewidth}\epsfig{file=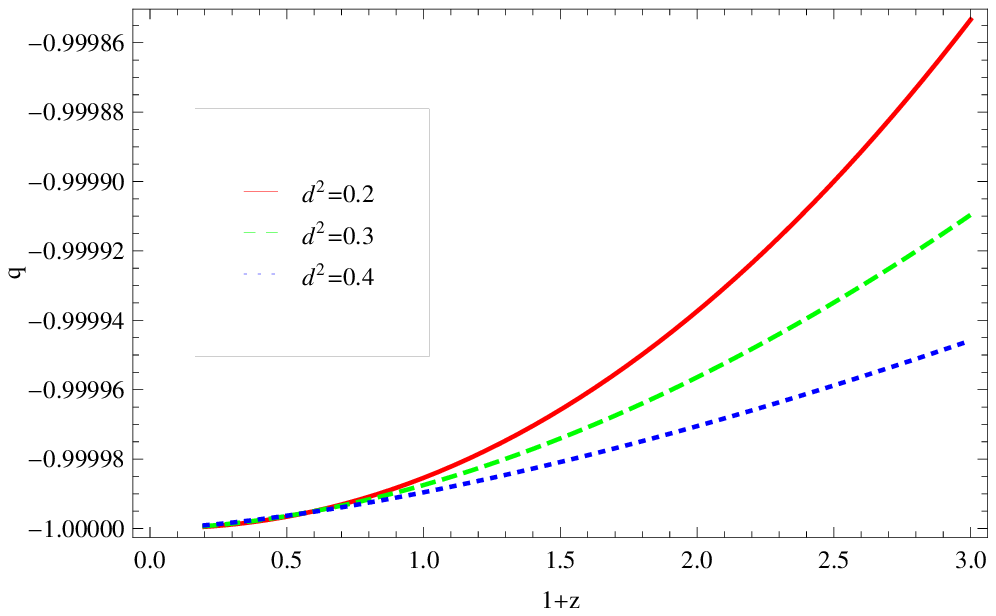,width=.50\linewidth}\\
\epsfig{file=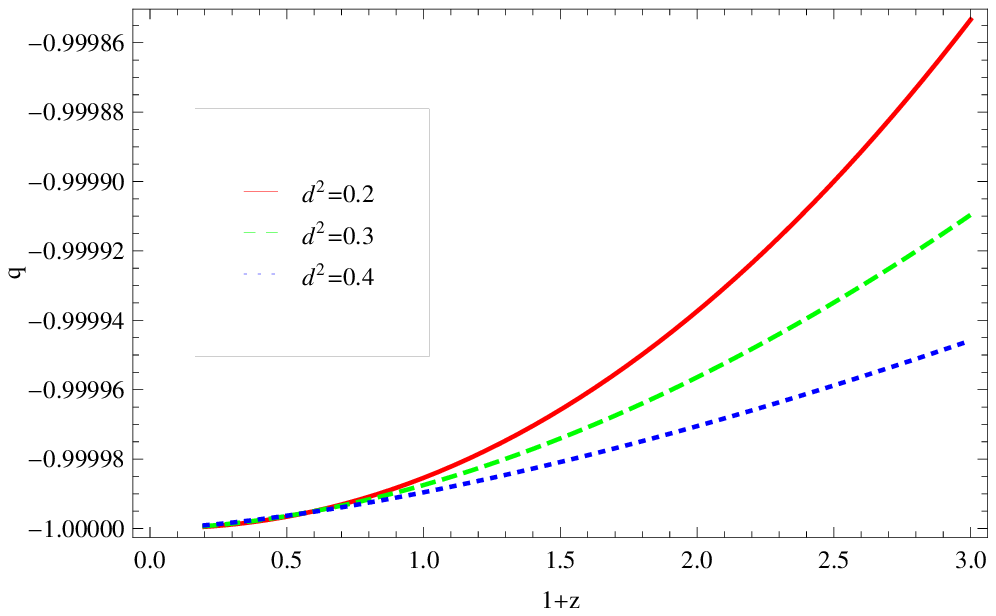,width=.50\linewidth}\caption{Plots
of $q$ versus $1+z$ for GGPDE model in DGP with u=1 (upper left
panel), $u=-1$ (upper right panel), $u=-2$ (lower panel),
respectively.}
\end{figure}
However, the EoS parameter can be obtained by using Eqs. (\ref{5})
and (\ref{30})
\begin{eqnarray}\nonumber
\omega_{D}&=&-1-\frac{b^{2}\rho_{m_{0}}a^{-3(1-b^{2})}}{(\alpha H
+\beta H^{2})^{u}}-\frac{u}{3H}\big(\frac{\alpha+2\beta H}{\alpha
H+\beta
H^{2}}\big)\bigg(-\frac{1}{H^{2}}\\\label{31}&\times&\big(\frac{(1-b^{2})\rho_{m_{0}}a^{-3(1-b^
{2})}}{2-2\epsilon\sqrt{\Omega_{r}}-\frac{u}{3H}(\alpha{H}+\beta{H^
{2}})^{u-1}(\alpha+2\beta{H})}\big)\bigg).
\end{eqnarray}
\begin{figure} \centering
\epsfig{file=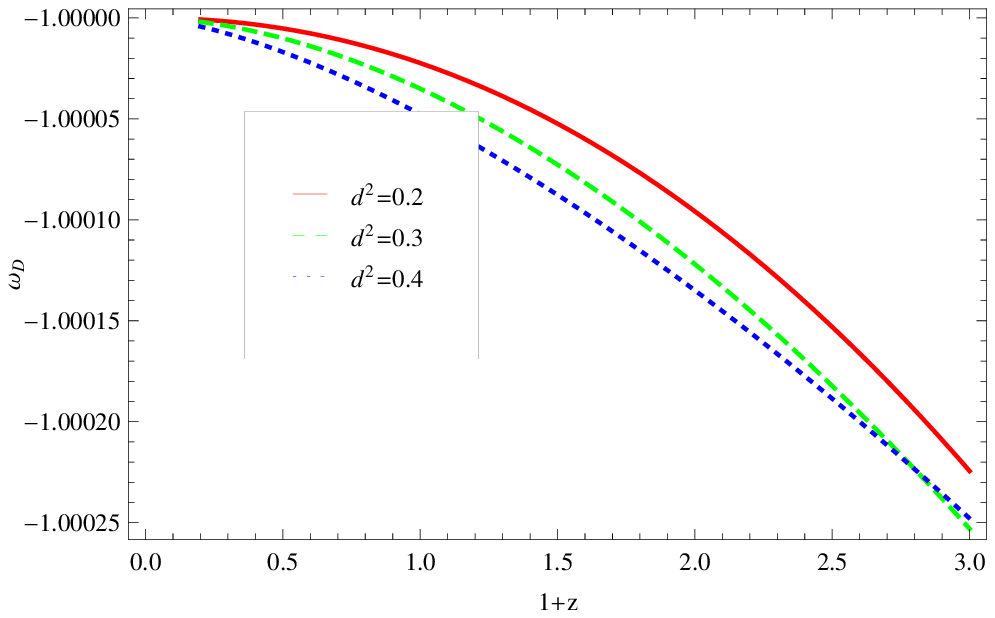,width=.50\linewidth}\epsfig{file=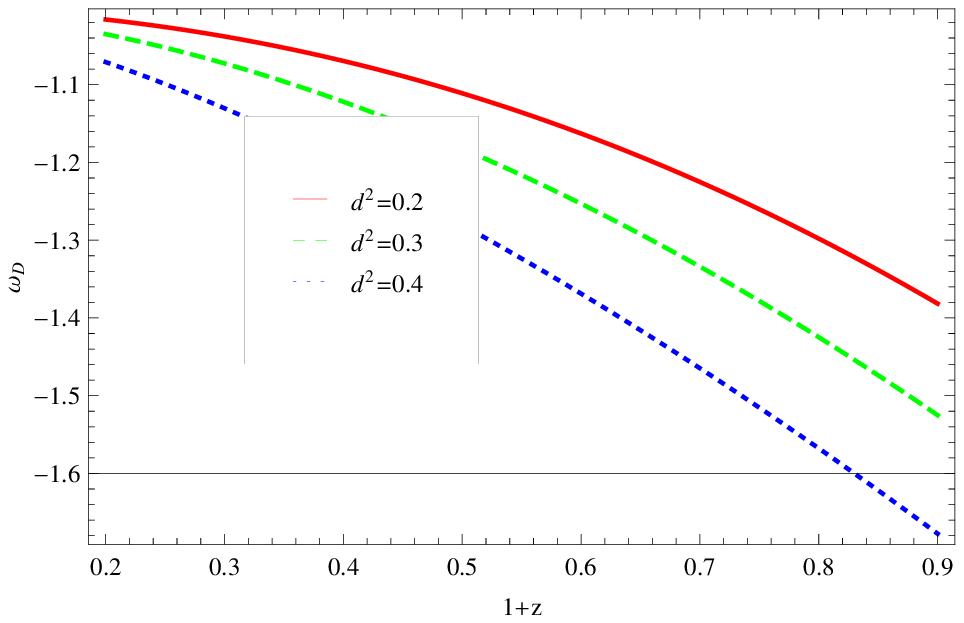,width=.50\linewidth}\\
\epsfig{file=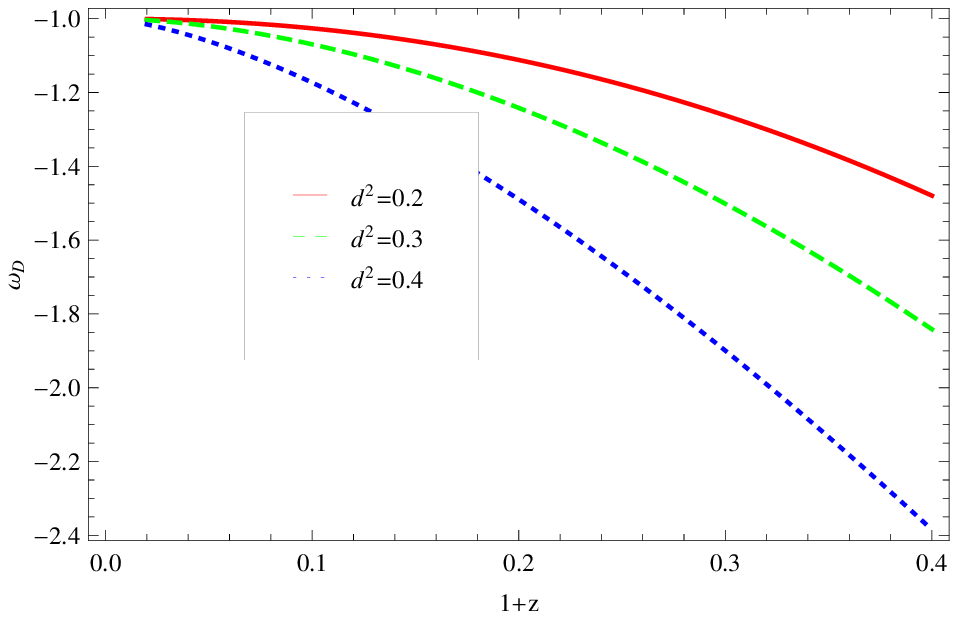,width=.50\linewidth}\caption{Plots of
$\omega_{\vartheta}$ versus $1+z$ for GGPDE model in DGP with u=1
(upper left panel), $u=-1$ (upper right panel), $u=-2$ (lower
panel), respectively.}
\end{figure}

The derivative of Eq.(\ref{30}) w.r.t. $x$ leads to
\begin{eqnarray}\nonumber
{\omega_{D}}'&=&3b^{2}(1-b^{2})\rho_{m_{0}}a^{-3(1-b^{2})}\big(\alpha
H+\beta
H^{2}\big)^{-u}+ub^{2}\rho_{m_{0}}a^{-3(1-b^{2})}\\\nonumber&\times&H^{-1}\big(\alpha
H+\beta H^{2}\big)^{-u-1}(\alpha+2\beta
H)\bigg(-(1-b^{2})\rho_{m_{0}}a^{-3(1-b^
{2})}\\\nonumber&\times&\big(2-2\epsilon\sqrt{\Omega_{r}}-\frac{u}{3H}(\alpha{H}+\beta{H^
{2}})^{u-1}(\alpha+2\beta
H)\big)^{-1}\bigg)-\frac{u}{3H^{2}}\\\nonumber&\times&(\frac{\alpha+2\beta
H}{\alpha H+\beta
H^{2}})\bigg(\bigg(3(1-b^{2})^{2}\rho_{m_{0}}a^{-3(1-b^
{2})}H\big(2-2\epsilon\sqrt{\Omega_{r}}-\frac{u}{3H}\\\nonumber&\times&(\alpha{H}+\beta{H^
{2}})^{u-1}(\alpha+2\beta
H)\big)^{-1}\bigg)+\bigg((1-b^{2})\rho_{m_{0}}a^{-3(1-b^
{2})}\\\nonumber&\times&\big(2-2\epsilon\sqrt{\Omega_{r}}-\frac{u}{3H}(\alpha{H}+\beta{H^
{2}})^{u-1}(\alpha+2\beta
H)\big)^{-2}\bigg)H^{-1}\\\nonumber&\times&\big(-(1-b^{2})\rho_{m_{0}}a^{-3(1-b^
{2})}\big(2-2\epsilon\sqrt{\Omega_{r}}-\frac{u}{3H}(\alpha{H}+\beta{H^
{2}})^{u-1}\\\nonumber&\times&(\alpha+2\beta
H)\big)^{-1}\big)\bigg(2\epsilon\sqrt{\Omega_{r}}-\frac{u(u-1)}{3}(\alpha{H}+\beta{H^
{2}})^{u-2}(\alpha\\\nonumber&+&2\beta
H)^{2}+\frac{u\alpha}{3H}(\alpha{H}+\beta{H^
{2}})^{u-1}\bigg)\bigg)+\frac{u}{3H^{2}}(2\alpha^{2}+
4H^{2}\beta^{2}\\\nonumber&+&5H\alpha\beta)(\alpha H+\beta
H^{2})^{-2}\bigg(-(1-b^{2})\rho_{m_{0}}a^{-3(1-b^
{2})}\big(2-2\\\label{32}&\times&\epsilon\sqrt{\Omega_{r}}-\frac{u}{3H}(\alpha{H}+\beta{H^
{2}})^{u-1}(\alpha+2\beta H)\big)^{-1}\bigg)^{2}.
\end{eqnarray}
\begin{figure} \centering
\epsfig{file=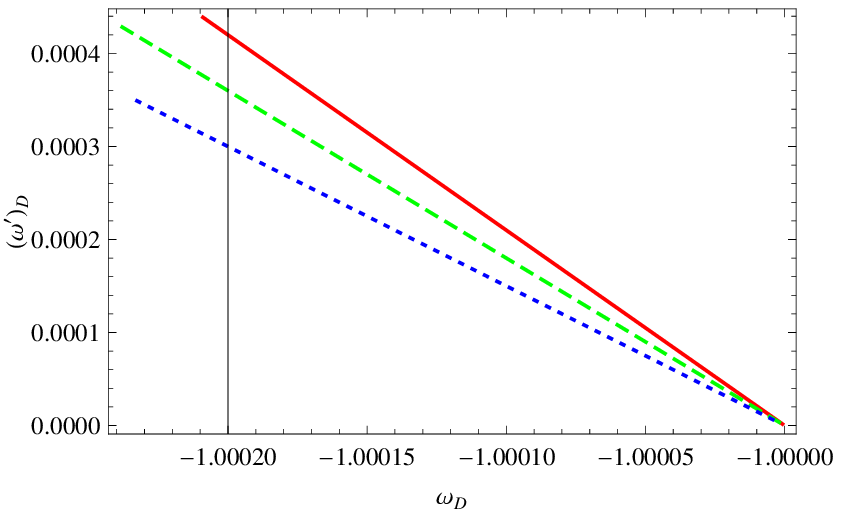,width=.50\linewidth}\epsfig{file=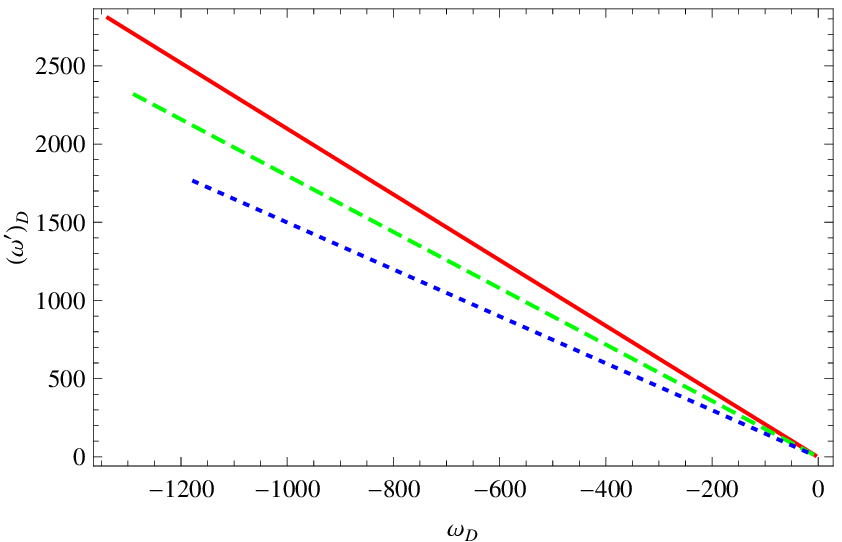,width=.50\linewidth}\\
\epsfig{file=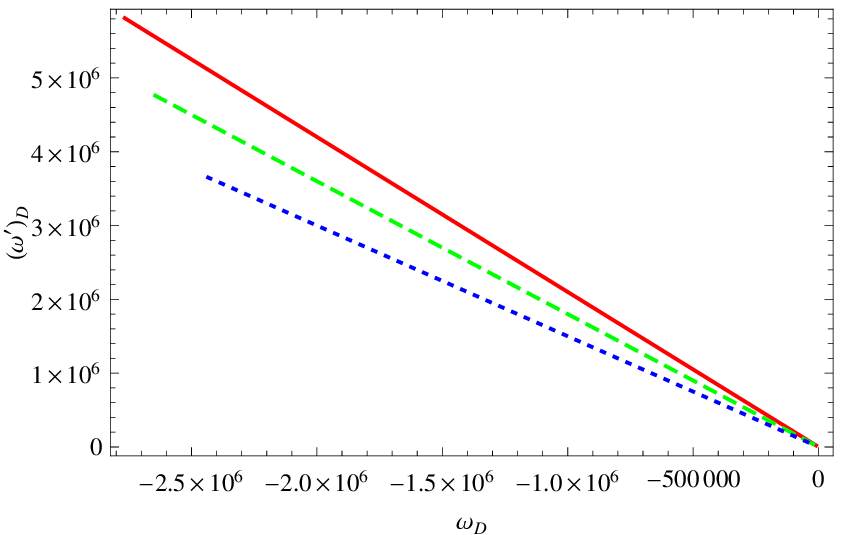,width=.50\linewidth}\caption{Plots of
$\omega_{D}-\omega'_{D}$  GGPDE model in DGP with u=1 (upper left
panel), $u=-1$ (upper right panel), $u=-2$ (lower panel),
respectively.}
\end{figure}
After solving the Eqs.(\ref{14}), (\ref{15}) and (\ref{28}), we can
obtain the statefinder parameter as
\begin{eqnarray}\nonumber
r&=&1-\bigg(\frac{1}{H^{2}}\big(\frac{(1-b^{2})\rho_{m_{0}}a^{-3(1-b^
{2})}}{2-2\epsilon\sqrt{\Omega_{r}}-\frac{u}{3H}(\alpha{H}+\beta{H^
{2}})^{u-1}(\alpha+2\beta{H})}\big)\bigg)\\\nonumber&\times&\bigg(3+
(1-b^{2})\rho_{m_{0}}a^{-3(1-b^{2})}\big(2-
2\epsilon\sqrt{\Omega_{r}}-\frac{u}{3H}(\alpha{H}+\beta{H^{2}})^{u-1}\\\nonumber&\times&
(\alpha+2\beta{H})\big)^{-2}\big(2-2\epsilon\sqrt{\Omega_{r}}-\frac{u
(u-1)}{3H}(\alpha{H}+\beta{H^{2}})^{u-2}\\\nonumber&\times&(\alpha+2\beta{H})^{2}+\frac{u
\alpha}{3H}(\alpha H+\beta H^{2})^{u-1}\big)\bigg)+\frac{3}{H^{2}}
\rho_{m_{0}}a^{-3(1-b^{2})}\\\label{33}&\times&(1-b^{2})\bigg(2-2\epsilon\sqrt{\Omega_{r}}
-\frac{u}{3H}(\alpha{H}+\beta{H^{2}})^{u-1}(\alpha+2\beta{H})\bigg).
\end{eqnarray}

\begin{eqnarray}\nonumber
s&=&\frac{1}{3}\bigg(-\frac{3}{2}+\frac{1}{H^{2}}\big(\frac{(1-b^{2})\rho_{m_{0}}a^{-3(1-b^
{2})}}{2-2\epsilon\sqrt{\Omega_{r}}-\frac{u}{3H}(\alpha{H}+\beta{H^
{2}})^{u-1}(\alpha+2\beta{H})}\big)\bigg)^{-1}\\\nonumber&\times&\bigg
(\frac{1}{H^{2}}\big(\frac{(1-b^{2})\rho_{m_{0}}a^{-3(1-b^
{2})}}{2-2\epsilon\sqrt{\Omega_{r}}-\frac{u}{3H}(\alpha{H}+\beta{H^
{2}})^{u-1}(\alpha+2\beta{H})}\big)\big(3+
(1-b^{2})\\\nonumber&\times&\rho_{m_{0}}a^{-3(1-b^{2})}\big(2-
2\epsilon\sqrt{\Omega_{r}}-\frac{u}{3H}(\alpha{H}+\beta{H^{2}})^{u-1}
(\alpha+2\beta{H})\big)^{-2}\\\nonumber&\times&\big(2-2\epsilon\sqrt{\Omega_{r}}-\frac{u
(u-1)}{3H}(\alpha{H}+\beta{H^{2}})^{u-2}(\alpha+2\beta{H})^{2}+\frac{u
\alpha}{3H}\\\nonumber&\times&(\alpha H+\beta
H^{2})^{u-1}\big)\big)+\frac{3}{H^{2}}
(1-b^{2})\rho_{m_{0}}a^{-3(1-b^{2})}\big(2-2\epsilon\sqrt{\Omega_{r}}
\\\label{34}&-&-\frac{u}{3H}(\alpha{H}+\beta{H^{2}})^{u-1}(\alpha+2\beta{H})\big)\bigg).
\end{eqnarray}
\begin{figure} \centering
\epsfig{file=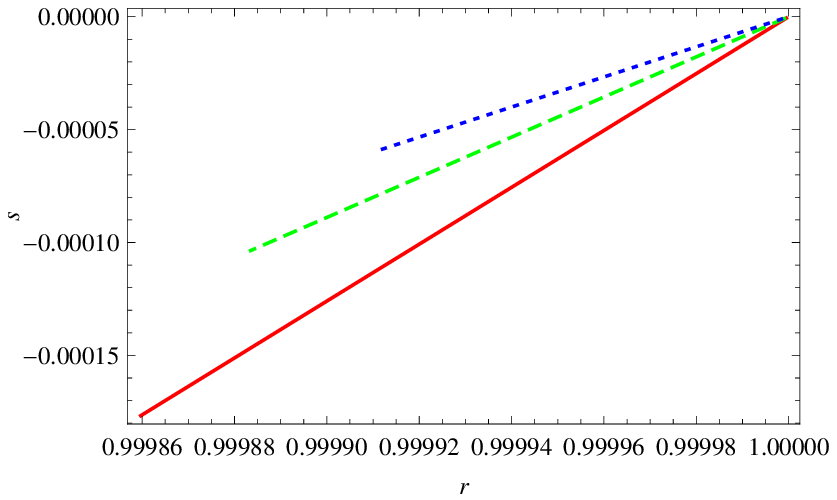,width=.50\linewidth}\epsfig{file=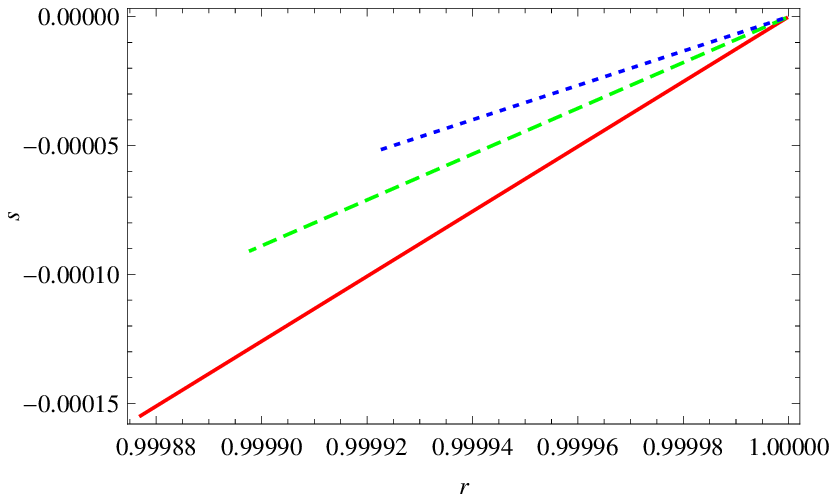,width=.50\linewidth}\\
\epsfig{file=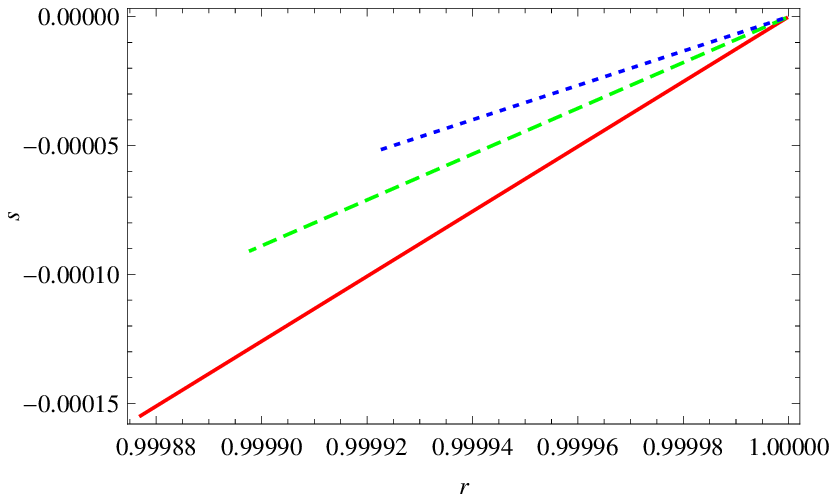,width=.50\linewidth}\caption{Plots
of $r-s$ for GGPDE model in DGP with u=1 (upper left panel), $u=-1$
(upper right panel), $u=-2$ (lower panel), respectively.}
\end{figure}

The deceleration parameter indicates the cosmic acceleration because
it lies within the range ($-1,0$) for all values of $u$ (Figure
\textbf{5}). For this case, EoS parameter represents the
phantom-like behavior of the universe for all values of $u=1, -1,
-2$ (Figure \textbf{6}). We also developed $\omega_{D}-\omega'_{D}$
plane for this case for $u=1,-1,-2$ as shown in Figure \textbf{7}.
It can be seen that this plane corresponding to thawing region as
well as to $\Lambda$CDM limit for all values of $u$. The $r-s$
planes for $u=1, -1, -2$ also approaches to $\Lambda$CDM limit for
all values of $u$ (Figure \textbf{8}).

\section{Conclusion}

In the present paper, we have investigated the cosmological
implications by assuming two interacting DE models such as PDE with
GO cutoff and GGPDE. We have explored various cosmological
parameters as well as planes for these two DE models. The results
are illustrated as follows: The EoS parameter have depicted the
phantom-like behavior for $u=1$ (left panel of Figure \textbf{1})
while exhibits quintessence-like behavior for $u=-1,-2$ (right and
lower panels of Figure \textbf{1}) for PDE with GO cutoff. However,
the phantom-like behavior of the universe for all values of $u=1,
-1, -2$ (Figure \textbf{6}) for GGPDE. The trajectories of
deceleration parameter have also indicated the accelerated expansion
of the universe in both DE models (Figures \textbf{2} and
\textbf{5}). The $\omega_{D}-\omega'_{D}$ plane corresponding to
$u=1$ lies in the thawing region, while lie in the freezing region
for $u=-1, -2$ for PDE with GO cutoff (Figure \textbf{3}). For
GGPDE, it can be seen that this plane corresponding to thawing
region as well as to $\Lambda$CDM limit for all values of $u$
(Figure \textbf{7}). For PDE with GO cutoff, the $r-s$ planes for
$u=1, -1, -2$ have shown in Figure \textbf{4} which behave like
chaplygin gas model. However, the $\Lambda$CDM limit is also
achieved for $u=-1$ case. On the other hand, the $r-s$ planes for
$u=1, -1, -2$ also approaches to $\Lambda$CDM limit for all values
of $u$. Finally, it is remarked that all the cosmological parameters
in the present scenario shows compatibility with the current
well-known observational data \cite{Planck}-\cite{wmap}.


\begin{thebibliography}{43}

\bibitem{auna} A. G. Riess et al., Astron. J.  {\bf 116}, 1009 (1998)

\bibitem{aunb} S. Perlmutter et al., Astrophys. J. {\bf 517}, 565 (1999)

\bibitem{aunc} P. de Bernardis et al., Nature {\bf 404}, 955 (2000)

\bibitem{aund} S. Hanany et al., Astrophys. J. {545}, L5 (2000).

\bibitem{adeuxa} P. J. E. Peebles and B. Ratra, Rev. Mod. Phys. {\bf 75}, 559 (2003)

\bibitem{adeuxb} T. Padmanabhan, Phys. Repts. {\bf 380}, 235 (2003)

\bibitem{d1} E. Komatsu et al (WMAP Collaboration) Astrophys. J. Suppl. {\bf 180}, 330-376,
(2009).

\bibitem{2s} Kamenshchik, A.Y., Moschella, U. and Pasquier, V.: Phys. Lett. B
\textbf{511}(2001)265.

\bibitem{3s} Hsu, S.D.H.: Phys. Lett. B \textbf{594}(2004)13.

\bibitem{4s} Li, M.: Phys. Lett. B \textbf{603}(2004)1.

\bibitem{5s} Cai, R.G.: Phys. Lett. B \textbf{660}(2008)113.

\bibitem{6s} Karami, K., Ghaffari, S. and Fehri, J.: Eur. Phys. J. C
\textbf{64}(2009)85.

\bibitem{N5} Wei, H.: Class. Quantum Grav. \textbf{29}, 175008
(2012).

\bibitem{N6} Sharif, M. and Jawad, A.: Eur. Phys. J. C \textbf{73}, 2382
(2013).
\bibitem{N7} Sharif, M. and Jawad, A.: Eur. Phys. J. C \textbf{73}, 2600
(2013).

\bibitem{st1} J.~Amor\'os, J.~de Haro and S.~D.~Odintsov, Physical Review D 87, {\bf 104037} (2013).

\bibitem{st2} E. V. Linder, Phys.Rev. D {\bf 81}, 127301 (2010) [Erratum-ibid. D 82, 109902 (2010)]

\bibitem{st3} M. Jamil, D. Momeni and R. Myrzakulov, Eur. Phys. J. C {\bf 72} (2012)
2267.

\bibitem{st4} R. Myrzakulov, Entropy {\bf 14} (2012) 1627.

\bibitem{st5} I.G.Salako, M.E.Rodrigues, A.V.Kpadonou, M.
J.S.Houndjo and J.Tossa: JCAP { \bf 060}, 1475-7516 (2013).

\bibitem{st6} M. E. Rodrigues, I. G. Salako, M. J. S. Houndjo, J. Tossa Int. J. Mod. Phys. D { \bf 23}, 1450004
(2014).


\bibitem{ma1} E. H. Baffou, A. V. Kpadonou, M. E. Rodrigues, M. J. S. Houndjo,
and J. Tossa Astrophys.Space Sci {\bf 355}, 2197 (2014).

\bibitem{ma2} M. J. S. Houndjo, Int. J. Mod. Phys. D. {\bf 21}, 1250003 (2012).

\bibitem{mj1}  S.~'i.~Nojiri and S.~D.~Odintsov, Phys.\ Lett.\ B {\bf 631}, 1 (2005). S. Nojiri, S. D. Odintsov, A.
Toporensky, P. Tretyakov, arXiv:0912.2488.

\bibitem{mj2} K. Bamba, S.
D. Odintsov, L. Sebastiani, S. Zerbini: arXiv:0911.4390.

\bibitem{mj3} K. Bamba, C.-Q. Geng, S. Nojiri, S. D. Odintsov: arXiv:0909.4397.

\bibitem{mj4} M.E. Rodrigues, M.J.S. Houndjo, D. Momeni, R. Myrzakulov:
arXiv:1212.4488.

\bibitem{mj5}M. J. S. Houndjo, M. E. Rodrigues, D. Momeni, R. Myrzakulov,
arXiv:1301.4642.

\bibitem{DGP} G. R. Dvali, G. Gabadadze, and M.Porrati, Phys. Lett. B 485 (2000) 208.

\bibitem{Deffayet} C. Deffayet, Phys. Lett. B 502 (2001) 199.

\bibitem{Deffayet1} C. Deffayet, G.R. Dvali, G. Gabadadze, Phys. Rev. D 65
(2002) 044023.

\bibitem{Koyama} K. Koyama Gen. Rel. Grav. 40 (2008) 421.

\bibitem{Hirano} K. Hirano and Z. Komiya Gen. Rel. Grav. 42 (2010) 2751,
[arXiv:0912.4950 [astroph.CO]].

\bibitem{Dvali} G. Dvali and M.S. Turner arXiv:astro-ph/0301510.

\bibitem{Koyama1} K. Koyama, Class. Quant. Grav. 24, R231 (2007)
[arXiv:0709.2399 [hep-th]].

\bibitem{Lue} A. Lue and G. D. Starkman, Phys. Rev. D
70, 101501 (2004) [arXiv:astro-ph/0408246].

\bibitem{Lazkoz} R. Lazkoz, R. Maartens and E. Majerotto, Phys. Rev. D 74,
083510 (2006) [arXiv:astro-ph/0605701].

\bibitem{Mariam} M. Bouhmadi-Lopez, JCAP 0911:011 (2009).

\bibitem{MariamChimento} M. Bouhmadi-Lopez and L. Chimento,
Phys.Rev.D82:103506 (2010).

\bibitem{28} Luty, M.A. et al.: JHEEP \textbf{09}(2003)029.

\bibitem{29} Goobar, A. et al.: Phys. Lett. B \textbf{642}(2006)432;
Majerotto, E. et al.: Phys. Rev. D \textbf{74}(2006)023004.

\bibitem{30} Wang, Y. et al.: Phys. Rev. D \textbf{82}(2010)043503.

\bibitem{31} Dvali, J.: New. J. Phys. \textbf{8}(2006)326; Koyoma,
K.: Class. Quant. Grav.  \textbf{24}(2007)231-253.

\bibitem{DGP1} Aguilera, Y.: Eur. Phys. J. C \textbf{74}(2014)3172.

\bibitem{DGP2} Ghaffari, S., Dehghani, H. and Sheykhi, A.:
Phys. Rev. D \textbf{89}(2014)123009.

\bibitem{DGP3} Ghaffari, S., Sheykhi, A. and Dehghani, H.:
Phys. Rev. D \textbf{\textbf{91}}(2015)023007.

\bibitem{RR14} t' Hooft, G.: \emph{Dimensional Reduction in Quantum Gravity}, gr-qc/9310026;
Susskind, L.: J. Math. Phys. \textbf{36}(1995)6377.

\bibitem{RR15} Cohen, A., Kaplan, D. and Nelson, A.: Phys. Rev. Lett. \textbf{82}(1999)4971.


\bibitem{20} Akama, A. Lect. Notes Phys.: \textbf{176}(1982)0001113.

%
%
%
%
%

\bibitem{30a} Jawad, A. and Shahzad, M. U..: Eur. Phys. J. C
\textbf{76}(2016)123.

\bibitem{30b} Sharif, M. and Jawad, A.: Int. J. Mod. Phys. D
\textbf{22}, 1350014 (2013).

\bibitem{30c} Jamil, M.: Eur. Phys. J. C
\textbf{62}(2009)325.

\bibitem{30d} Bhadra, J. and Debnath, U.: Eur. Phys. J. C
\textbf{72}(2012)1912.

\bibitem{shara} Sharif, M. and Jawad, A.: Eur. Phys. J. Plus \textbf{129}, 15
(2014).

\bibitem{sharb} Lobo, F.S.N.: Phys. Rev. D \textbf{71}(2005)124022.

\bibitem{sharc} Lobo, F.S.N.: Phys. Rev. D \textbf{71}(2005)084011.

\bibitem{shard} Sushkov, S.: Phys. Rev. D \textbf{71}(2005)043520.
%

%
\bibitem{9} Gonzalez, J.A. and Guzman, F.S.: Phys. Rev. D \textbf{79}(2009)121501.

\bibitem{10} Sun, C.Y.: Commun. Theor. Phys. \textbf{52}(2009)441.

\bibitem{11} Harada, T., Maeda, H. and Carr, B.J.: Phys. Rev. D \textbf{74}(2006)024024;
Akhoury, R., Gauthier, C.S. and Vikman, A.: JHEP 03(2009)082.

\bibitem{S1} Cai, Y-F., et al.: Phys. Reports \textbf{493}(2010)1.

\bibitem{S2} Saridakis, E.N.: Nucl. Phys. B {\bf 819}(2009)116.

\bibitem{S3} Gupta, G., Saridakis, E.N. and Sen, A.A.: Phys. Rev. D \textbf{79}(2009)123013.

\bibitem{S4} Setare, M.R. and Saridakis, E.N.: JCAP
\textbf{0903}(2009)002.

\bibitem{S4} Setare, M.R. and Saridakis, E.N.: Phys. Lett. \textbf{B}
671(2009)331.

\bibitem{S5} Saridakis, E.N., Gonzalez-Diaz, P.F. and Siguenza, C.L.: Class. Quant. Grav.
\textbf{26}(2009)165003.

\bibitem{S6} Saridakis, E.N.: Phys. Lett. B \textbf{676}(2009)7.

\bibitem{S7} Saridakis, E.N.: Phys. Lett. B \textbf{660}(2008)138.

\bibitem{S8} Saridakis, E.N.: Phys.Lett.B661:335-341,2008.

\bibitem{S9} Setare, M.R. and Saridakis, E.N.: Phys. Lett. B
\textbf{671}(2009)331.

\bibitem{N47} Sharif, M. and Rani, S.: J. Exp. Theor. Phys. (to appear,
2014).

\bibitem{N48} Chattopadhyay, S., Jawad, A., Momeni, D. and Myrzakulov,
R.:Astrophys. Space Sci. \textbf{353}, 279 (2014).

\bibitem{N49} Jawad, A.: Astrophys. Space Sci. \textbf{360}(2015)52.

\bibitem{RR17} Granda, L. and Oliveros, A.: Phys. Lett. B \textbf{669}(2008)275.


\bibitem{RR35} Yu, F. et al.: Phys. Lett. B \textbf{688}(2010)263.



\bibitem{RR36} Wang, Y. and Xu, L.: Phys. Rev. D \textbf{81}(2010)083523.









%
%
%
%
%
%

%
%
%
%
%
%
%
%
%
%
%
%
%
%
%
%
%
%
%
%
%
%
%
%
%
%
%

\bibitem{46} Granda, L.N. et al.: Phys. Lett. B \textbf{669}(2008)275.

\bibitem{75} Caldwell, R.R. and Linder, E.V.: Phys. Rev. Lett. \textbf{95}(2005)141301.


%
\bibitem{40} Sahni, V. et al.: JETP. Lett. \textbf{77}(2003)201.
%
\bibitem{41} Alam, U. et al.: Astron. Soc. \textbf{344}(2003)1057.
%
%
%
%
%

%
%
%
%
%
%
%
%
%
%
%
%
%
%
%
%
\bibitem{62} Sharif, M., Jawad, A.: Astrophy. Space. Sci.
\textbf{351}(2014)321.

\bibitem{Planck} Ade, P.A.R., et al.: Ade, P.A.R., et al.: A.A. \textbf{571}
(2014)A16.

\bibitem{Riess} Riess, A. G., et al.: Astrophys. J.\textbf{730}(2011)119.

\bibitem{wmap} Hinshaw, G.F. et al.: Astrophys. J. Suppl. \textbf{208}(2013)19.


\end{thebibliography}
\end{document}